**Tidal Response of Groundwater in a Leaky Aquifer – Application to Oklahoma**

Chi-Yuen Wang[1], Mai-Linh Doan[2], Lian Xue[1], Andrew J. Barbour[3]

[1.] Dept. of Earth and Planetary Science, University of California, Berkeley, CA 94720

[2.] Univ. Grenoble Alpes, Univ. Savoie Mont Blanc, CNRS, IRD, IFSTTAR, ISTerre, 38000 Grenoble, France

[3.] Earthquake Science Center, U.S. Geological Survey, Menlo Park, CA 94025

**Abstract**

Quantitative interpretation of the tidal response of water levels measured in wells has long been made either with a model for perfectly confined aquifers or with a model for purely unconfined aquifers. However, many aquifers may be neither totally confined nor purely unconfined at the frequencies of tidal loading but behave somewhere between the two end members. Here we present a more general model for the tidal response of groundwater in aquifers with both horizontal and vertical flow. The model has three independent parameters: the transmissivity ($T$) and storativity ($S$) of the aquifer, and the specific leakage ($K'/b'$) of the leaking aquitard, where $K'$ and $b'$ are the hydraulic conductivity and the thickness of the aquitard, respectively. If $T$ and $S$ are known independently, this model may be used to estimate aquitard leakage from the phase shift and amplitude ratio of water level in wells obtained from tidal analysis. We apply the model to interpret the tidal response of water level in a USGS deep monitoring well installed in the Arbuckle aquifer in Oklahoma, into which massive amount of wastewater co-produced from hydrocarbon exploration has been injected. The analysis shows that the Arbuckle aquifer is leaking significantly at this site. We suggest that the present method may be effectively and economically applied to monitor leakage in groundwater systems, which bears on the safety of water resources, the security of underground waste repositories, and the outflow of wastewater during deep injection and hydrocarbon extraction.

**1. Introduction**

The response of aquifers to applied loads, such as Earth tides and barometric pressure, have long been studied for the evaluation of aquifer properties [e.g., Hsieh, et al., 1987; Roeloffs, 1996; Allègre, et al., 2016; Xue, et al., 2016] and their changes after earthquakes [e.g., Elkhoury, et al., 2006; Doan, et al., 2006; Liao, et al., 2015; Zhang, et al., 2015]. Interpretations of such responses have been made with models either for perfectly confined aquifers or for purely unconfined aquifers. Most aquifers, however, behave somewhere between these two end members [e.g., Galloway and Rojstaczer, 1989]. The vertical impedance to flow across the boundary of a confined aquifer is not infinite, and the response of aquifers to applied load depends on the time scale. With applied loading at low frequencies, a confined aquifer may exchange flow across its boundaries; and at high frequencies, an unconfined aquifer may exhibit some 'confined' behaviors. Thus the analysis of aquifer response to applied loads may benefit from the inclusion of a frequency-dependent leakage.

A second motivation for inclusion of leakage in the study of aquifer response to applied loads comes from the coseismic response of water level to earthquakes. Studies have shown that permeability of aquifers may change after earthquakes probably due to seismic shaking that dislodges debris and/or multiphase droplets or bubbles from pre-existing fractures [Beresnev and Johnson, 1994; Brodsky et al., 2003; Beresnev et al. 2005; Elkhoury et a., 2006; Liu and Manga, 2009; Manga et al., 2012]. If so, one may expect enhanced permeability to occur not only in the horizontal direction but in all directions since the pre-existing fractures are likely to be randomly oriented. Furthermore, earthquake-enhanced vertical permeability has been invoked to explain coseismic increases in streamflow [e.g., Wang, et al., 2004a; Wang and Manga, 2015], eruption of geothermal water [e.g., Wang, et al., 2004b], changes in groundwater temperature [e.g., Wang, et al., 2012, 2013], coseismic changes in the tidal response of water levels [Liao et al., 2015; Wang et al., 2016], and migration of seismic swarms [e.g., Ingebritsen and Manning, 2010]. Thus the study of groundwater response to earthquakes may also benefit from the consideration of leakage in the system.

A third motivation for the inclusion of leakage in the study of aquifers is for monitoring the safety of groundwater resource and/or the security of underground repositories. While most aquifers are used as sources of fresh water, some aquifers are used for disposal of wastewater and other hazardous liquids. In either case, it is important to monitor if leakage occurs. For decades massive amounts of wastewater, coproduced from the extraction of oil and gas, have been injected into deep aquifers beneath the U.S. mid-continent [Frohlich, 2012; Ellsworth, 2013; Keranen, et al., 2013, 2014; Hornbach, et al., 2015; McGarr, et al., 2015; Walsh and Zoback, 2015; Weingarten, et al., 2015] and disposal activities continue to this day. Concerns arise if the injected fluids can migrate upward and contaminate shallow groundwater [Vidic, et al., 2013; U.S. Environmental Protection Agency, 2016]; even though such an event has not been documented [Darrah, et al., 2014], the issue remains contentious [Vengosh, et al., 2014]. While continuous monitoring of leakage may be advisable in such situations, traditional methods such as well tests, numerical simulation, and geochemical monitoring are costly and labor intensive – infeasible for continuous monitoring. Here we show that the analysis of the tidal response of water levels in wells provides an effective means for continuous monitoring of leakage in groundwater systems.

As noted earlier, the interpretation of the response of aquifers to Earth tides has been traditionally made either with a model for perfectly confined aquifer or with a model for purely unconfined aquifer. In this study we derive a new analytical solution for the response of groundwater in a leaky aquifer to Earth tides. We apply the model to analyze the tidal response of water level in a USGS deep monitoring well installed in the Arbuckle aquifer in Oklahoma, where massive injection of wastewater co-produced from hydrocarbon exploration is active.

## 2. Previous studies

The study of groundwater pumping in a leaky system has a long history. Analytical solutions for pumping/injection in a leaky, multilayered-aquifer system have been developed since early last century. *Hantush and Jacob* [1955] and *Hantush* [1960] considered steady state and transient flow through the aquitard. Solutions were extended to multilayered systems [*Hemker*, 1985; *Maas*, 1987a, b; *Cheng and Morohunfola*, 1993; *Hemker and Maas*, 1994; *Cheng*, 1994; *Veling and Maas*, 2009] and used to investigate pressure change in response to fluid injection or extraction in wells [Cihan, et al., 2011; Cardiff, et al., 2013; Sun, et al., 2015].

The study of groundwater response to the solid Earth tide is different from that of groundwater pumping. In the pumping studies, the driving force in a well is treated mathematically as a boundary condition, while in the study of groundwater response to Earth tide the driving force acts on every point of the groundwater system. Furthermore, while the study of groundwater pumping in a leaky system has a long history, the study of the response of a leaky groundwater system to the solid Earth tide is at its infancy, as described below.

The classical model of tidal response of groundwater in a confined aquifer by Hsieh, et al. [1987] exploits the phase shift caused by the time needed for groundwater in the aquifer to flow into and out of the well; it predicts the a negative phase shift of water level oscillation relative to the tidal strain. Another model is for unconfined aquifer with purely vertical flow [Roeloffs, 1996; Wang, 2000], which predicts an apparent positive phase shift of water level oscillation relative to the local tidal volumetric strain. This difference in the sign of phase shift predicted by the two models has been used in previous studies as the primary criterion for deciding if an aquifer is confined or unconfined and thus which of the above two models should be used in interpreting the tidal response [e.g., Elkhoury, et al., 2006; Liao, et al., 2015; Zhang, et al., 2015; Allègre, et al., 2016; Xue, et al., 2016]. However, as noted earlier, many aquifers may neither be totally confined nor purely unconfined at the frequencies of tidal loading, but behave somewhere between the two end members. Here we present a more general model with both horizontal and vertical flow for the response of groundwater to the solid Earth tide. We show that substantial leakage may occur when the phase shift is negative; thus negative phase shift in tidal response alone may not be a reliable criterion for deciding if an aquifer is confined.

## 3. Tidal Response of a Leaky Aquifer

Here we derive the response of the basic Hantush leaky aquifer to the solid Earth tide. The model consists of an aquifer confined above by a semi-confining aquitard that in turn is overlain by an unconfined aquifer (Figure 1). The model applies Darcy's law across the entire aquitard of thickness $b'$ and hydraulic conductivity $K'$ and implicitly assumes that the aquitard is incompressible and has zero storage. The analytical technique for tidal analysis presented below builds upon previous works [Hsieh, et al., 1987; Doan, et al., 2006] and extends to the Hantush leaky aquifer.

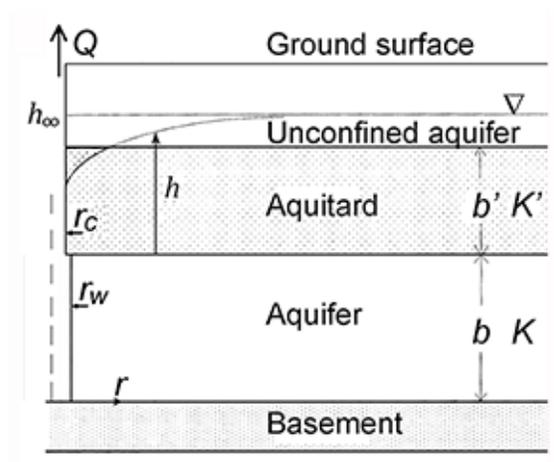

*Figure 1. The Hantush model for leaky aquifer. Vertical dashed line on the left shows the position of well axis located at r = 0. The thickness (b) and hydraulic conductivity (K) of the aquifer are related to the aquifer transmissivity by T = bK. The equivalent thickness and vertical hydraulic conductivity of the aquitard are b' and K', respectively.*

Assuming that the aquifer is laterally extensive and that flow through the semi-confining aquitard is vertical, the tide-induced groundwater flow in the leaky aquifer may be evaluated by solving the following equation:

$$T\left[\frac{\partial^2 h}{\partial r^2} + \frac{1}{r}\frac{\partial h}{\partial r}\right] - \frac{K'}{b'}h = S\left(\frac{\partial h}{\partial t} - \frac{BK_u}{\rho g}\frac{\partial \varepsilon}{\partial t}\right) \qquad (1)$$

where $h$ is the hydraulic head above a common reference (Figure 1), $r$ is the radial distance from the studied well, $T$ and $S$, respectively, are the transmissivity and storativity of the aquifer, $\varepsilon$ is the tidal oscillating volumetric strain of the aquifer, $B$ and $K_u$, respectively, are the Skempton's coefficient and the undrained bulk modulus of the aquifer, and $K'$ and $b'$, respectively, are the vertical hydraulic conductivity and the thickness of the aquitard. A list of the notations and their definitions is provided in the Supporting Information. The model in Equation (1) differs from the classical model [Hsieh, et al., 1987] in its inclusion of the vertical leakage, approximated by $-K'h/b'$ and treated as a volumetric source term, which, as noted earlier, implicitly assumes that the aquitard is incompressible with negligible storage and the flow across it is vertical [e.g., Lee, 1999]. These assumptions may be justifiable if leakage through the aquitard is controlled by narrow vertical cracks. The topmost unconfined aquifer has high hydraulic conductivity; thus its hydraulic head is likely to be hydrostatic [Hantush and Jacob, 1955; Lee, 1999].

The boundary conditions are

$$h(r,t) = h_\infty(t) \quad \text{at } r = \infty, \qquad (2)$$

$$h(r,t) = h_w(t) \quad \text{at } r = r_w, \text{ and} \qquad (3)$$

$$2\pi r_w T (\partial h/\partial r)_{r=r_w} = \pi r_c^2 (\partial h_w/\partial t) \qquad (4)$$

where $h_w = h_{w,o}e^{i\omega t}$ is the periodic water level in well with angular frequency $\omega$ and complex amplitude $h_{w,o}$, $r_w$ is the radius of the screened portion of the well, and $r_c$ is the inner radius of well casing in which water level fluctuates with tides (Figure 4).

Following Hsieh, et al. (1987), we use complex notation to facilitate the model development below. The solution is obtained by first deriving the response away from the well, $h_\infty$, and then modifying it by taking into account the effect of the well on aquifer response by using a flux condition at the well that accounts for wellbore storage. Let the disturbance in water level due to the well be expressed as

$$\Delta h(r,t) = h(r,t) - h_\infty(t) \tag{5}$$

where $h_\infty(t)$ is the hydraulic head away from the well (Figure 4a), which is a function of time only and may be evaluated by replacing $h$ by $h_\infty$ in equation (1):

$$-\frac{K'}{b'}h_\infty = S\frac{\partial h_\infty}{\partial t} - \frac{SBK_u}{\rho g}\frac{\partial \varepsilon}{\partial t}. \tag{6}$$

Since $h_\infty$ and $\varepsilon$ are both periodic with the same frequency $\omega$ we have

$$h_{\infty,o} = \frac{i\omega S}{i\omega S + K'/b'}\left(\frac{BK_u\varepsilon_o}{\rho g}\right). \tag{7}$$

where $h_{\infty,o}$ is the complex amplitude of $h_\infty$ and $\varepsilon_o$ the amplitude of $\varepsilon$. It is notable that leakage causes both the amplitude and the phase shift of $h_{\infty,o}$ to deviate from that of a perfectly confined aquifer and that $h_{\infty,o}$ reduces to that of a perfectly confined aquifer when $K' = 0$.

Replacing $h$ in equations (1) to (4) by $\Delta h + h_\infty$ and using equation (7) we have

$$T\left[\frac{\partial^2 \Delta h}{\partial r^2} + \frac{1}{r}\frac{\partial \Delta h}{\partial r}\right] - \frac{K'}{b'}\Delta h = S\frac{\partial \Delta h}{\partial t}. \tag{8}$$

Since the steady-state solution of equation (8) has the form $\Delta h = \Delta h_o(r)e^{i\omega t}$, the above equation may be reduced to an ordinary differential equation

$$T\left[\frac{d^2\Delta h_o}{dr^2} + \frac{1}{r}\frac{d\Delta h_o}{dr}\right] - \frac{K'}{b'}\Delta h_o = i\omega S \Delta h_o. \tag{9}$$

with the boundary conditions

$$\Delta h_o(r \to \infty) = 0, \tag{10}$$

$$\Delta h_o(r = r_w) = h_{w,o} - h_{\infty,o} = h_{w,o} - \frac{i\omega S}{i\omega S + K'/b'}\left(\frac{BK_u\varepsilon_o}{\rho g}\right), \tag{11}$$

$$2\pi r_w T \frac{d\Delta h_o}{dr}\bigg|_{r=r_w} = i\omega\pi r_c^2 h_{w,o}. \tag{12}$$

The solution to equation (9) is $\Delta h_o = C_I I_o(\beta r) + C_K K_o(\beta r)$, where $I_o$ and $K_o$ are, respectively, the modified Bessel functions of the first and second kind and the zeroth order, and

$$\beta = \left(\frac{K'}{Tb'} + \frac{i\omega S}{T}\right)^{1/2}. \tag{13}$$

The boundary condition (equation 10) asserts that $C_I = 0$; thus $\Delta h_o = C_K K_o(\beta r)$. Solving for $C_K$ with equation (12) and recalling $\frac{dK_o(r)}{dr} = -K_1(r)$, where $K_1$ is the modified Bessel function of the second kind and the first order, we have

$$C_K = -\frac{i\omega r_c^2 h_{w,o}}{2T\beta r_w K_1(\beta r)}.$$

Thus,

$$\Delta h_o = -\frac{i\omega r_c^2 h_{w,o} K_o(\beta r)}{2T\beta r_w K_1(\beta r)}. \tag{14}$$

Inserting equation (14) into equation (11) we finally have,

$$h_{w,o} = \frac{i\omega S}{(i\omega S + K'/b')\xi}\left(\frac{BK_u\varepsilon_o}{\rho g}\right) \tag{15}$$

where

$$\xi = 1 + \left(\frac{r_c}{r_w}\right)^2 \frac{i\omega r_w}{2T\beta} \frac{K_0(\beta r_w)}{K_1(\beta r_w)}. \tag{16}$$

An independent derivation of equation (15) using Laplace transform is given in the Supporting Information. The solution has three independent parameters, *T* and *S* for the aquifer and *K'/b'* for the semi-confining aquitard. We define the amplitude ratio of the tidal response as

$$A = \left| h_{wo} / \left(\frac{BK_u \varepsilon_o}{\rho g}\right) \right|, \tag{17}$$

and the phase shift is defined as

$$\eta = \arg\left[ h_{wo} / \left(\frac{BK_u \varepsilon_o}{\rho g}\right) \right], \tag{18}$$

where arg(*z*) is the argument of the complex number *z*. Figure 2 shows the amplitude ratio and the phase shift of the M2 (semidiurnal lunar) tide against *K'/b'* at selected values of *T* and *S*. We focus on interpreting the observed tidal phases because the amplitude ratio requires knowledge on $K_u$ and *B*, i.e., equation (18), that are often unknown or have large uncertainties. We will comment later on the use of the amplitude ratio for checking the consistency of the model with measurements.

Several aspects of Figure 2 are worthy of notice. First, increasing leakage (*K'/b'*) causes the phase to increase and the amplitude ratio to decrease. Second, leakage can be significant when the phase is negative if $T < 10^{-4}$ m$^2$/s. For example, the curve for $T = 10^{-6}$ m$^2$/s and $S = 0.01$ in Figure 2 predicts a specific leakage of $K'/b' \sim 10^{-6}$ s$^{-1}$ for a phase shift of -20 degree. If *b'* (aquitard thickness) is ~100 m, the corresponding *K'* (vertical conductivity of the aquitard) is $\sim 10^{-4}$ m/s, which is similar to that of a common aquifer (e.g., Ingebritsen, et al., 2006). Thus observing a negative phase shift in the tidal response is not necessarily an indication of perfect confinement: adequate interpretation of tidal response requires the consideration of aquifer leakage. Third, at small *K'/b'* and constant *T* and *S*, phase shift stays nearly constant until *K'/b'* increases beyond a certain threshold that depends upon *T* and *S*; above that threshold the phase shift increases (or becomes less negative) significantly with *K'/b'*. For example, the curve for $T = 10^{-6}$ m²/s and $S = 0.01$ shows that phase shift is nearly constant for $K'/b' < 10^{-7}$ s$^{-1}$; significant increases of phase shift occur only when *K'/b'* exceeds $10^{-7}$ s$^{-1}$. Thus *K'/b'* may be estimated only above this threshold. Finally, at $T > 10^{-4}$ m²/s, the curves for different *T* and a given *S* overlap and appear as a single curve on the diagram. This is because the phase difference between tidal response of water level in well and pore pressure in the aquifer approaches zero at such high transmissivity [e.g., Doan, et al., 2006]. Naturally, the amplitude ratio is further constrained by the logger resolution; it cannot be smaller than the resolution of the logger.

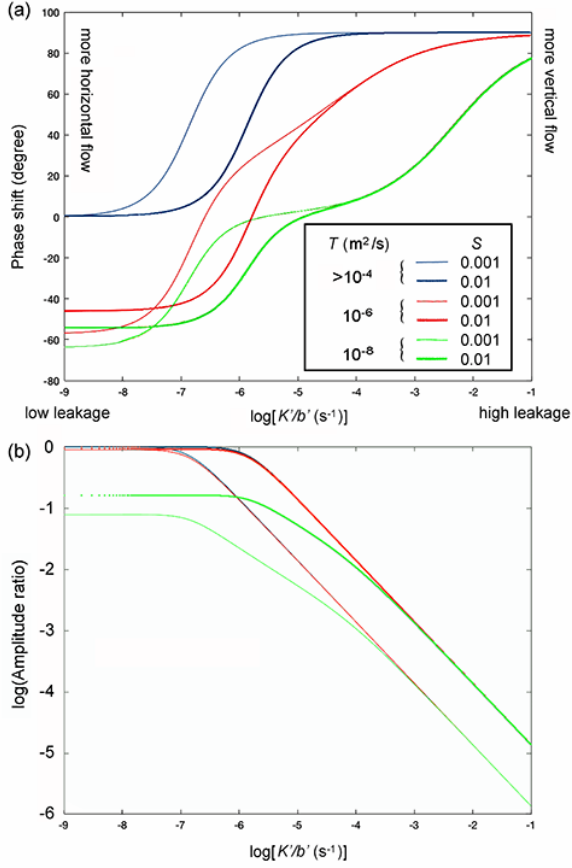

*Figure 2. (a) Phase shift of water level response to the M2 (semidiurnal lunar) tide, plotted against the logarithm of the specific leakage (K'/b') for different T and S, with $r_c$ = 3.65 cm and $r_w$ = 11 cm. Negative values indicate local phase lag. (b) Logarithm of the ratio of the amplitude of water level response to that of the volumetric strain, plotted against the logarithm of K'/b' for different T and S.*

Verification of equation (15) against published analysis cannot be made because no such analysis is available. Partial verification of the solution may be made by letting $K' = 0$. Equation (15) then reduces to

$$h_{wo} = \frac{BK_u\varepsilon_o}{\rho g} \frac{1}{\xi}\bigg|_{K'=0}, \tag{19}$$

which is identical to the classical solution for a perfectly confined aquifer [Hsieh, et al., 1987; Doan, et al., 2006]. Figure 3 further shows that the predicted phase shift and amplitude ratio for the O1 (diurnal lunar) and M2 tides by equation (15) at $K' = 0$ match seamlessly with those predicted by perfectly confined aquifer.

On the other hand, equation (15) cannot be reduced to the classical solution for a purely unconfined aquifer at $T = 0$ because, while the classical unconfined aquifer model is specifically that of a half space, the leaky aquifer model developed here is for an aquifer of finite thickness and confined below. More discussion on this point is given in the Supporting Information.

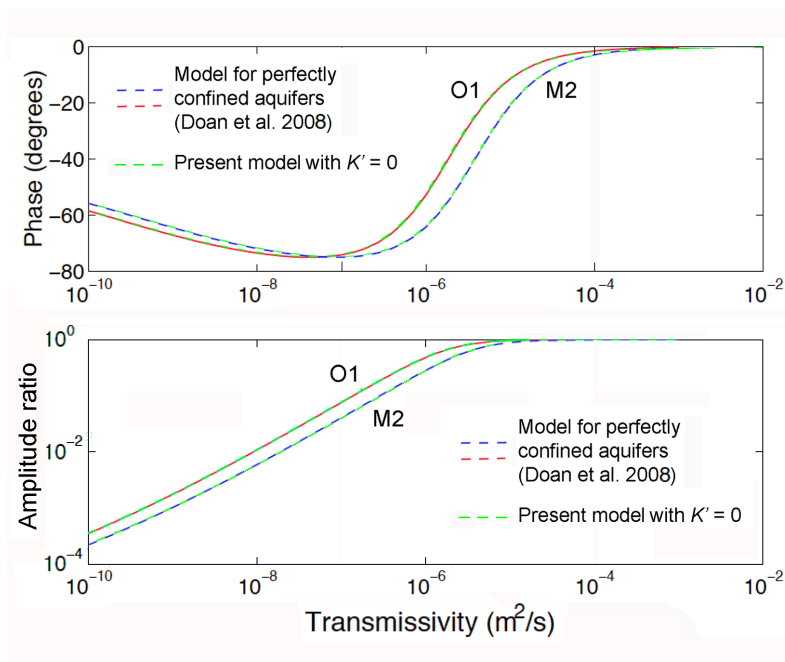

Figure 3. (a) Phase shift and (b) amplitude ratio of water level response to the O1 and M2 tides predicted by the present model with $K' = 0$, compared with that predicted by a perfectly confined aquifer [Hiesh, et al., 1987; Doan, et al., 2006].

## 4. Application of the leaky aquifer model to the Arbuckle aquifer, Oklahoma

For decades, massive amounts of wastewater have been injected into the deeply buried part of the Arbuckle aquifer of Oklahoma, but volumes have increased substantially in the last decade. With it has followed dramatic increases in seismicity rate [Ellsworth, 2013; McGarr, et al., 2015; Walsh and Zoback, 2015; Weingarten, et al., 2015], including several M≥5 earthquakes [Keranen, et al., 2013, 2014; McNamara, et al., 2015; Yeck, Hayes, et al., 2016; Yeck, Weingarten, et al., 2016; Barbour, et al., 2017] (see Figure 4 for locations of disposal wells and epicenters of major earthquakes in 2016). In April 2017 the U.S. Geological Survey installed a pressure gauge in a deep monitoring well in the Arbuckle aquifer in northeastern Oklahoma (see Figure 4 for well location with respect to injection wells and Table 1 for detailed well information), measuring water levels continuously at a rate of one sample per minute.

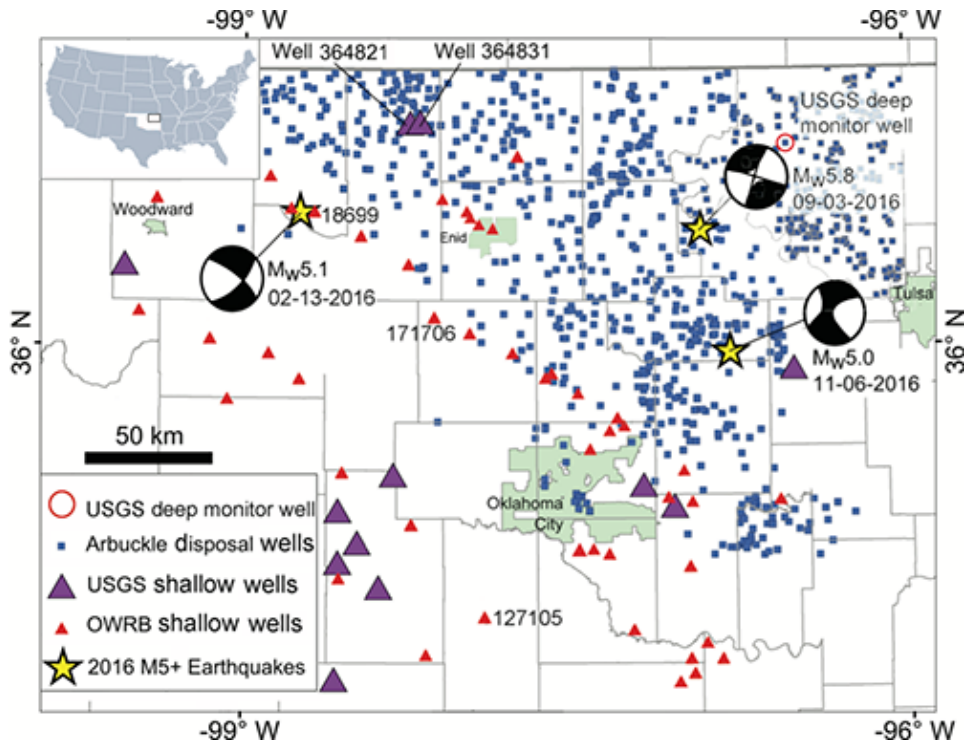

*Figure 4. Location of the USGS Oklahoma deep monitoring well (red circle on top right corner), together with the locations of Arbuckle disposal wells, shallow USGS and OWRB (Oklahoma Water Resources Board) monitoring wells, and the epicenters of large earthquakes in 2016. Inset map on upper left of diagram shows the study area (small rectangle) in the State of Oklahoma (white polygon).*

The Arbuckle formation is a thick deposit of laterally extensive, dominantly Late Cambrian to Early Ordovician limestone and dolomite over a Proterozoic to Early Cambrian igneous basement in the U.S. mid-continent [Johnson, 2008]. During the Late Carboniferous Period the aquifer was deformed, uplifted, eroded and exposed in south-central Oklahoma. Beneath north-central Oklahoma, however, this aquifer is deeply buried and confined by younger formations [Johnson, 2008]. The well log in Figure 5 shows that the Arbuckle aquifer near the USGS deep monitoring well is confined by a sequence of sedimentary strata including a basal shale, sandstones and carbonate rocks, which in turn is overlain by an unconfined aquifer of younger sediments. This stratigraphic sequence corresponds closely with the conceptual model of Hantush leaky aquifer described above (Figure 1), with the sequence of basal shale, sandstone and carbonate rocks above the Arbuckle aquifer (Figure 5) representing the semi-confining aquitard.

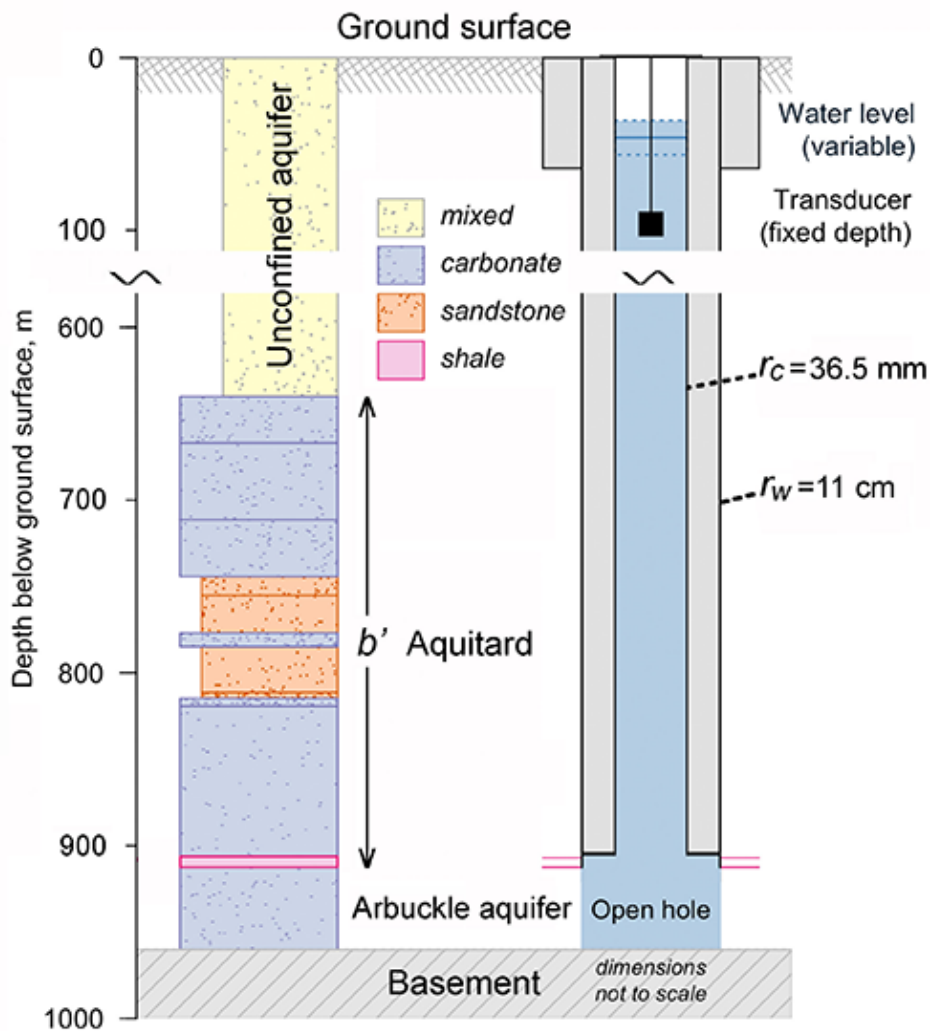

Figure 5. *Simplified completion diagram of the USGS Oklahoma deep monitoring well. The Arbuckle aquifer is the lowest sedimentary rock that lies above the basement. Comparing with the conceptual model in Figure 1, the sequence of sedimentary rocks between the Arbuckle aquifer and the topmost unconfined aquifer, which includes a basal shale, sandstone and carbonate, makes up the aquitard. The unconfined aquifer consists of unconsolidated sediments.*

**4.1 Tidal response of water level in the USGS monitoring well**

Water level in the USGS Oklahoma deep monitoring well is measured with a pressure sensor LevelTROLL 500 manufactured by In Situ (https://in-situ.com/wp-content/uploads/2014/11/SS_LevelTROLL_Spec_Sheet_Dec2017.pdf). It is a vented, piezo-resistive transducer made of titanium, with a nominal accuracy of 0.05% full scale. The signal is digitized at the surface at a sampling rate of 1/4 Hz, low-pass filtered to 1 min sampling, and sent by satellite telemetry to the U.S. National Water Information System.

We use the code Baytap08 [Tamura, et al., 1991] for extracting tidal signals from the data. The method is based on Bayesian statistics with the prior knowledge that the time series comprises tidal components with known periods, and a drift that includes long-period and secular changes. Figure 6 shows the time series of raw data for water level above the mean sea level in the USGS Oklahoma deep monitoring well. Figures 6b to 6c show, respectively, the drift that was removed and the remaining tides used in the analysis. There is no meteorological station at or very near the well; thus the barometric effect on water level is not corrected and we focus on the response to the M2 tide because it is less affected by changes in the barometric pressure. The effect of ocean tides at the USGS well is small because of the large distance of the well from the coasts; calculations using SPOTL (a software for modeling the response to ocean-tide loading [Agnew, 2012]) show that the ocean-tide effect is ~5% of that of the solid Earth tide.

The period of the M2 tide (0.5175 day) is close to that of the S2 (semidiurnal solar) tide (0.5000 day); thus spectral leakage between the S2 and the M2 tides can pose challenges [Allègre, et al., 2016]. We choose a window size of 29.5036 days, the minimum time window needed to separate the frequencies of the semidiurnal M2 and S2 tides [Allègre, et al., 2016; Xue, et al., 2016]. Figure 6d shows the time-varying phase shift of water level response to the M2 and S2 tides, referenced to the local volumetric strain tides. Negative phase shift indicates phase lag and positive indicates phase advance. The root-mean-square errors for the determinations are ~0.3°, on average. Large and variable changes in the phase shift of the S2 tide are probably due to changes in barometric pressure and temperature; whereas, the phase shift of the M2 tide is positive and stable at ~12.5° throughout the studied period, demonstrating that the two tides are well separated in the analysis. The amplitude response of water level to the M2 tide is also stable at ~4.5 cm (Figure 6e), while that of the S2 tide shows much less stability.

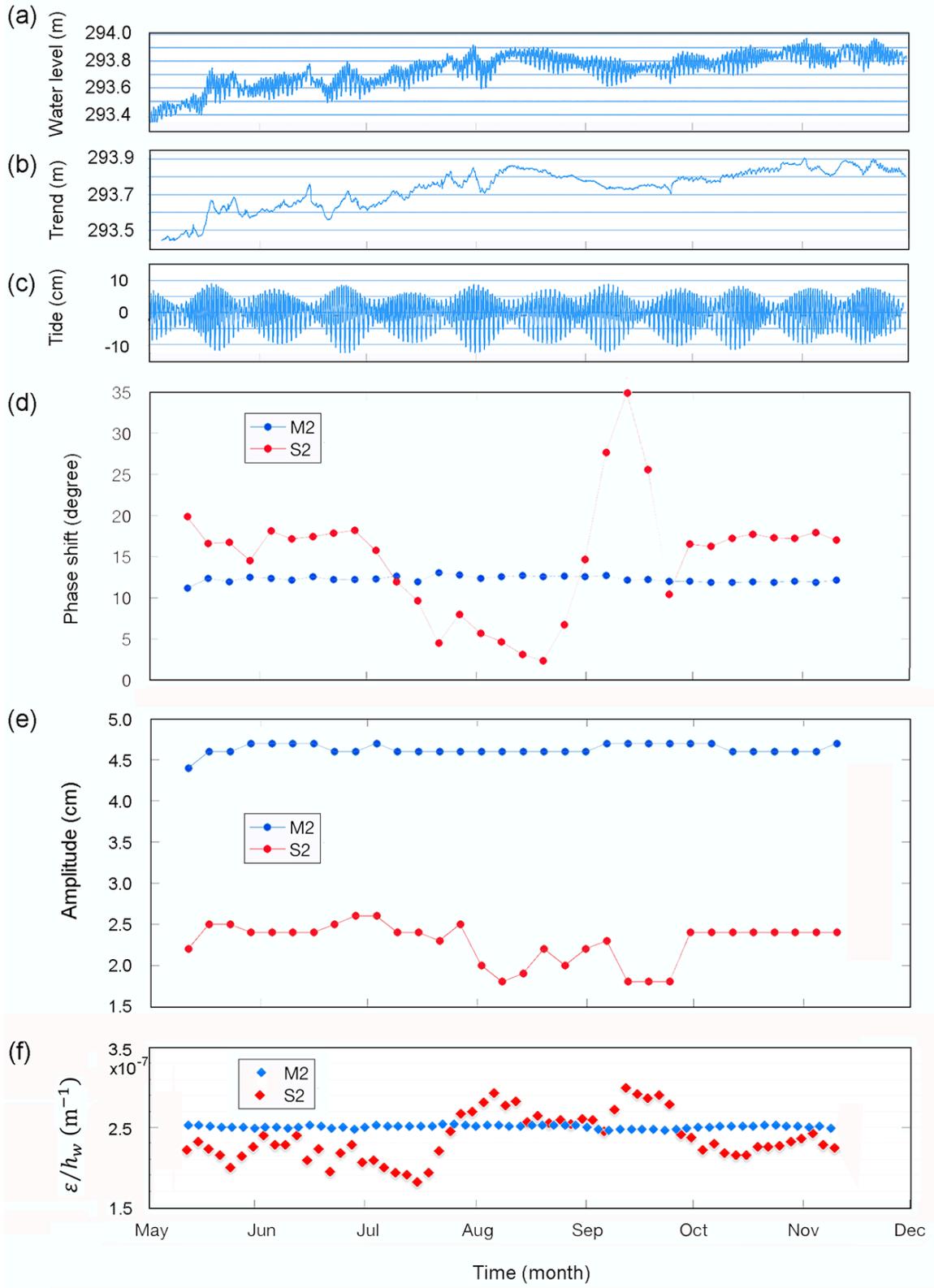

*Figure 6. Time series of (a) raw data for water level above the mean sea level in the USGS Oklahoma deep monitoring well, (b) drift that was removed, (c) remaining tides in water level used in the analysis, (d) phase shift of water level response to the M2 and S2 tides referenced to the local tidal volumetric strain, (e) amplitude of water level response to the M2 and S2 tides, and (f) response of $\varepsilon_o/h_{w,o}$ to the M2 and S2 tides, where $\varepsilon_o$ is the amplitude of the volumetric strain converted from surface strain computed in Baytap08.*

### 4.2 Interpretation of the tidal response

As noted earlier, both geologic studies [e.g., Johnson, 2008] and well logs (e.g., Figure 5) show that the Arbuckle aquifer is confined – an ideal target for massive injection of wastewater. Thus, the positive phase shift of water level from the above analysis (Figure 6d) was unexpected and suggests that the confining units of the Arbuckle near the USGS deep well may be leaking. In this section we use the model for a leaky aquifer derived in section 3 to interpret the tidal response of the Arbuckle aquifer with data from the USGS Oklahoma deep monitoring well (Figure 6). Table 1 lists the other hydrogeological parameters for the Arbuckle aquifer needed to interpret the measured phase shift from tidal analysis. In particular, the permeability ($k$) measured on small samples from the Arbuckle aquifer [Morgan and Murray, 2015] shows a range from $2\times10^{-14}$ to $3\times10^{-12}$ m², and the specific storage ($S_s$) obtained from tidal analysis of groundwater level in south-central Oklahoma [Rahi and Halihan, 2009] shows a range from $5.4\times10^{-8}$ to $5.6\times10^{-7}$ m⁻¹. For the inference of the groundwater leakage we use the median-to-maximal range of the measured permeability because small-scale matrix permeability most likely represents the lower end of permeability for the Arbuckle aquifer. These parameters are related to $T$ and $S$ by the following relations, respectively: $k = (\mu/\rho g b)T$ and $S_s = S/b$, where $g$ is the gravitational acceleration, $\rho$ and $\mu$ are, respectively, the density and viscosity of pore fluid in the Arbuckle aquifer. Given the aquifer thickness of 48 m in the USGS well, the range of $S_s$ corresponds to a range of $S$ from $2.6 \times 10^{-6}$ to $2.7 \times 10^{-5}$ and the range of $k$ corresponds to a range of $T$ from $9.6 \times 10^{-6}$ to $1.4 \times 10^{-3}$ m²/s. More data for $T$ and $S$ are needed for better interpretation of the measured phase shift at the USGS deep monitoring well.

Figure 7a shows the model curves (equation 15) for phase shift to the M2 tide versus log ($K'/b'$) for the range of $T$ and $S$ in Table 1. For a given value of $S$, the curves lie close together for the realistic range of $T$; however, for a given value of $T$, the curves for the realistic range of $S$ (Table 1) lie apart. Figure 7b shows the model curves (equation 14) for amplitude ratio versus log($K'/b'$) for the range of $T$ and $S$ in Table 1; here the curves for the range of $T$ overlap at a given $S$.

The phase shift of 12.5° for the water level response to the M2 tide in the USGS well (Figure 2d), represented by a purple horizontal line in Figure 7a, intersects the model curves at $K'/b'$ of $10^{-10}$ to $10^{-9}$ s$^{-1}$ for $S = 2.6 \times 10^{-6}$ to $2.7 \times 10^{-5}$. Given the thickness of 277 m for the semi-confining aquitard (Table 1), these values correspond to $K' \sim 3 \times 10^{-8}$ to $3 \times 10^{-7}$ m/s, respectively. As shown in the next section, this result provides the basic evidence that the Arbuckle aquifer is leaking.

Table 1. Parameters of the USGS Oklahoma deep well and the hydrogeological parameters used in estimating the vertical conductivity of the leaking aquitard.

| Parameters | Symbol | Values | References |
| --- | --- | --- | --- |
| Well location and elevation | | 36.7269N, 96.5317W 340.16 m above sea level | USGS website[&] |
| Well depth | | 960 m beneath surface | This study |
| Well radius | $r_w$ | 11 cm | This study |
| Casing radius | $r_c$ | 3.65 cm | This study |
| Thickness of aquitard | $b'$ | 277 m | Figure 5 |
| Thickness of aquifer | $b$ | 48 m | Figure 5 |
| Permeability* | $k$ | $2 \times 10^{-14}$ to $3 \times 10^{-12}$ m$^2$ | Morgan & Murray, 2015 |
| Transmissivity[@] | $T$ | $9.6 \times 10^{-6}$ to $1.4 \times 10^{-3}$ m$^2$/s | Calculated from $k$ |
| Specific storage | $S_s$ | $5.4 \times 10^{-8}$ to $5.6 \times 10^{-7}$ m$^{-1}$ | Rahi & Halihan, 2009 |
| Storativity[#] | $S$ | $2.6 \times 10^{-6}$ to $2.7 \times 10^{-5}$ | Calculated from $S_s$ |

[&] https://waterdata.usgs.gov/nwis/uv/?site_no=364337096315401

* Permeability was measured on the outcrop surface and core measurements. We use the median to maximum range of measured values because small-scale matrix permeability represents the lower end of permeability for the Arbuckle aquifer [Morgan and Murray, 2015].

[@]Transmissivity is calculated from permeability using the relationship $T=b(\rho g k/\mu)$, where $\rho$ and $\mu$ are, respectively, the density and viscosity of pore fluid in the

Arbuckle aquifer. As explained in the text, groundwater in the Arbuckle aquifer near the USGS well is similar to freshwater; thus we take $\rho$ = 1000 Kg/m³ and $\mu$ = 0.001 Pa s in the calculation of $T$ from $k$.

#Storativity $S$ is calculated from specific storage $S_s$ [Rahi and Halihan, 2009] using the relationship $S=bS_s$.

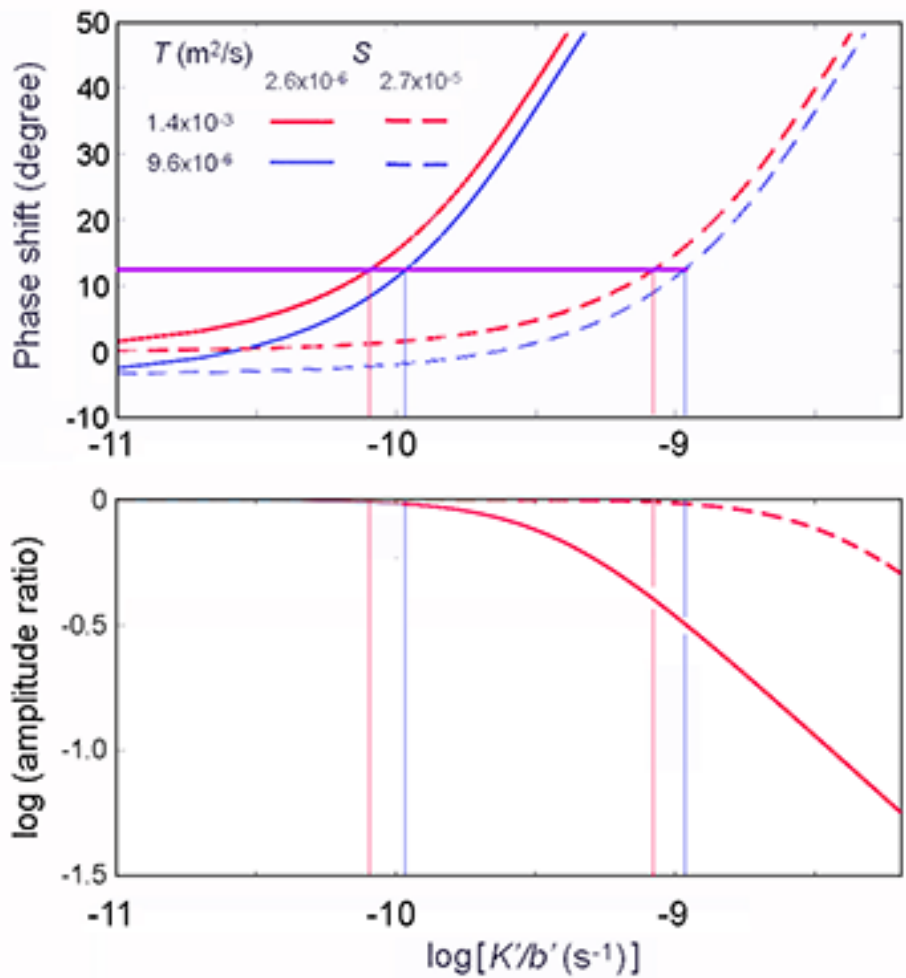

*Figure 7. (a) Calculated phase shift of water-level response to M2 tide as a function of K'/b' (colored curves) with predetermined values of T and S (Table 1) compared with observed phase shift in the USGS well (purple horizontal line). Intersections of the*

*horizontal line with colored curves give K'/b' of the aquitard. (b) Calculated amplitude ratio of water-level response to M2 tide as a function of K'/b' (colored curves). Curves calculated with different values of T but the same S overlap on this diagram. Vertical lines correspond to the estimated K'/b' from (a), which intersect the respective colored curves at amplitude ratios of ~1.*

## 5. Discussion

Several aspects of the above analysis are discussed below, including the assessment and verification of the leakage of the Arbuckle aquifer, the estimate of the leakage rate, the electrical conductivity and water level in the USGS deep well, and the criteria for separating the leakage effect on tidal response from that of enhanced horizontal permeability.

### 5.1. Assessment on Leakage of the Arbuckle Aquifer

We may examine the hydraulic integrity of the aquitard above the Arbuckle aquifer by comparing the above estimated $K'$ with that of a hypothetical, intact aquitard consisting of the same sequence of layers as shown in the well log (Figure 5), each assigned with a representative hydraulic conductivity according to its lithology. The average vertical hydraulic conductivity of a sequence of horizontal layers may be estimated from the harmonic mean of the vertical hydraulic conductivity of the individual layers [Ingebritsen, et al., 2006], i.e., $K' = b' / \sum_i (b_i/K_i)$, where $K_i$ is the vertical hydraulic conductivity of the $i^{th}$ layer in the aquitard. This relation shows that the average vertical conductivity of the horizontal layers in the aquitard is controlled by the layer with the lowest conductivity. Table 2 lists the thickness of each individual layer in the aquitard and its representative hydraulic conductivity, assigned according to the intact rock of the lithology of the layer. The calculated average vertical hydraulic conductivity of the hypothetical aquitard is ~$5 \times 10^{-12}$ m/s and is controlled by the 6 m thick intact shale. This conductivity is many orders of magnitude lower than that estimated from tidal analysis ($10^{-8}$ to $10^{-7}$ m/s). The basal shale of the aquitard would need to have a vertical hydraulic conductivity many orders of magnitude greater than that of intact (unfractured) shale in order to raise the calculated average vertical conductivity to the same order as that determined from tidal analysis. We therefore conclude that the basal shale above the Arbuckle aquifer near the USGS well is leaking, due perhaps to the presence of conductive fractures.

Table 2. Thickness and assumed permeability of rocks in calculating the harmonic mean of vertical permeability of a hypothetical, hydraulically intact aquitard

| Rock layer | Thickness (m) | Vertical hydraulic conductivity (m/s) | Reference |
|---|---|---|---|
| Carbonate | 106 | $10^{-6}$ | Morgan & Murray, 2015 |
| Sandstone | 31 | $10^{-8}$ | Wang, 2000 |
| Carbonate | 9 | $10^{-6}$ | Morgan & Murray, 2015 |
| Sandstone | 28 | $10^{-8}$ | Wang, 2000 |
| Carbonate | 92 | $10^{-6}$ | Morgan & Murray, 2015 |
| Shale | 6 | $10^{-13}$ | Wang, 2000 (for Piere shale) |

**5.2. Verification of the leakage assessment**

Although there is no independent evidence near the USGS well to corroborate the result of the above assessment that the Arbuckle is leaking, hydrogeological simulations of groundwater flow in south central Oklahoma [Christenson, et al., 2011] show that significant vertical conductivity of the layers above the Arbuckle aquifer is required to fit observational data.

We may also test the consistency of the above result against existing laboratory measurements of rock properties. Figure 6b shows that the amplitude ratios of the tidal response of water level in the USGS deep well is ~1 for the range of $K'/b'$ estimated above and the relevant $T$ and $S$ (Table 1). Thus $h_{w,o} \approx BK_u\varepsilon_o/\rho g$, where $\varepsilon_o$ is the amplitude of the oscillating volumetric strain in response to the M2 tide. From tidal analysis we have $h_{w,o} \approx 0.045$ m (Figure 6e). With $\varepsilon_o$ converted from the theoretical surface strain using a Poisson ratio of 0.25, we have $\varepsilon_o/h_{w,o} \approx 2.5\times10^{-7}$ m$^{-1}$ (Figure 6f). Thus $BK_u$ ~40 GPa, which falls close to the upper bound of the range of published values for consolidated rocks from laboratory measurements [Table C1 in Wang (2000)].

Kroll, et al. [2017] also estimated the poroelastic parameters for the Arbuckle formation based on the analysis of the coseismic response of water levels in some deep wells to large (M ≥ 5) induced earthquakes in Oklahoma. The approach was based on the assumption that the coseismic water-level response was caused by static volumetric strain estimated from a dislocation model with a set of earthquake source parameters [Kroll, et al., 2017]. Wang and Barbour [2017] compiled and analyzed the existing published measurements of coseismic volumetric strain and showed that most measured coseismic volumetric strains disagree with that calculated from the dislocation model. Thus additional mechanisms may play a role in the coseismic change of volumetric strain.

### 5.3. Estimate the Leakage Rate

Given the value of *K'/b'* from tidal analysis, we may estimate the rate of leakage across the aquitard near the USGS deep well. Figure 6a shows that the average hydraulic head of the Arbuckle aquifer was ~293.6 m above sea level (asl) during the time of this study, or ~46.6 m beneath the ground surface, given the ground elevation at the well (Table 1). Although there is no shallow well data near the USGS well, the groundwater table in Oklahoma is mostly near the surface [Wang, et al., 2017] and pore pressure in the unconfined aquifer is likely to be hydrostatic, as noted earlier. Thus the rate of leakage is given by *K'h/b'* where $h \sim 48$ m. Given the range of *K'/b'* estimated from the above tidal analysis (i.e., $10^{-10}$ to $10^{-9}$ s$^{-1}$), we estimate a downward leakage of groundwater from the unconfined aquifer into the Arbuckle at a rate of $4.8 \times 10^{-9}$ to $4.8 \times 10^{-8}$ m/s, or 0.15 to 1.5 m/yr, near the USGS deep monitoring well.

### 5.4. Electrical conductivity and water level in the USGS deep well

The specific electrical conductivity of groundwater in the USGS deep monitoring well was measured in April of 2017 and lies between 0.005 and 0.05 S/m. Since the USGS deep monitoring well is cased from the surface to the top of the Arbuckle aquifer (Figure 5), water in the well comes solely from the Arbuckle aquifer. This measured specific electrical conductivity of the groundwater is within the range for freshwater, which is unexpected because it is at least an order of magnitude lower than the specific electrical conductivity of the flow-back fluids injected at the well before the USGS operation started (e.g., Edwards, et al., 2011; Li, et al., 2014). Two mechanisms may have operated to dilute the concentration of the injected fluid. First, as shown in the above discussion, downward leakage of groundwater from the unconfined aquifer to the Arbuckle aquifer may have occurred. Groundwater in the unconfined aquifer is recharged from fresh surface water; thus the downward leakage of groundwater may have diluted the injected fluid and lowered its electrical conductivity. Available records show that a total of 926 bbl (~150 m³) were injected in October and November of 2014, and otherwise none. An approximate estimate of the degree of dilution of the injected fluid by the downward groundwater leakage is given below. Tidal oscillations of water level in the well induce lateral groundwater flow between the well and the aquifer, causing advective mixing of the injected fluid with groundwater. The lateral dimension of this mixing may be approximated by the characteristic diffusion length $\sqrt{\tau T/S}$, where $\tau$ = 0.5175 day is the period of the M2 tide. Given $T$ and $S$ for the Arbuckle aquifer (Table 1), we estimate that the lateral mixing occurs in an area extending ~100 m around the USGS deep well. The average concentration of the injected water around the USGS well is thus ~150/[$\pi$(100)²] m³/m² ~ 5x10⁻³ m³/m². At a rate of 0.15 to 1.5 m/year, the downward leakage from the unconfined aquifer to the Arbuckle aquifer between the end of injection (November 2014) and the time of conductivity measurement (April, 2017) would have added an amount of freshwater of ~0.4 to 4 m³/m² to the aquifer. Thus the concentration of the injected wastewater in the aquifer at the time of conductivity measurement would have been diluted by freshwater by a ratio of $10^{-2}$ to $10^{-3}$, which may explain the measured conductivity in the USGS deep well.

A second mechanism is lateral dispersion of the injected fluid by the local groundwater flow in the Arbuckle aquifer. However, since there is no available information about the velocity of groundwater flow near the USGS deep monitoring well, quantitative test of these hypotheses is difficult and beyond the scope of this study.

Water level in the Arbuckle aquifer shows significant secular change with time (Figure 6b). Could this change be related to the downward leakage from the unconfined aquifer to the Arbuckle? The rate of downward leakage of freshwater from the unconfined aquifer, as estimated earlier, is in the range between 0.15 and 1.5 m/yr. The average rate of water level increase in the USGS deep well was ~1.5 m/yr between May and August, 2017 (Figure 2b), similar to the upper bound of the estimated downward leakage. Between August and December, 2017, however, the average rate of water level increase is nearly zero (Figure 6b). Furthermore, the timing of the change in the rate of water-level change does not correspond to that of the change in the injection rate in the well. Thus further testing of this hypothesis is needed with longer time monitoring of water level change in the well.Another possible leak is into the igneous basement beneath the Arbuckle aquifer, which is not discussed in this study. The contact between the Arbuckle and the basement is an unconformity and likely to be hydraulically conductive. Most induced earthquakes in Oklahoma occur in the basement [e.g., Schoenball and Ellsworth, 2017], suggesting that some injected fluid must have leaked into the basement [Zhang, et al., 2013; Barbour, et al., 2017]. The size of this leak is difficult to estimate but is likely to be small in view of the small porosity of the basement rocks.

**5.5. Separating the leakage effect on tidal response from that of enhanced horizontal permeability**

As noted earlier, considerable leakage of a confined aquifer may occur at negative phase shift. Thus it may be challenging to separate the effect of enhanced horizontal permeability on the tidal response of a confined aquifer from the effect of increased vertical leakage. For the tidal response of a confined aquifer, the increase in phase shift due to an enhancement of the horizontal permeability is associated with an increase in the amplitude ratio (Hsieh, et al., 1987; Doan, et al., 2006). On the other hand, the increase of phase shift due to increased vertical leakage is associated with a decrease in the amplitude ratio (Figures 2 and 6). Thus changes in both the phase and amplitude ratio are needed in order to differentiate between the effect of enhanced horizontal permeability and that of increased vertical leakage. If permeability also increases in the horizontal direction, there will be an additional increase in phase shift, but the increase in amplitude ratio will offset the decrease due to increased vertical permeability. In this case, a large increase in phase shift may be associated with a reduced amplitude ratio.

Finally, we call attention to the simple nature of the model. As noted earlier, the leakage term $-K'h/b'$ in equation (1) is a linear approximation of Darcy's law and the formulation implicitly assumes that the aquitard is incompressible with negligible storage. Furthermore, the effects of many geologic complexities such as local topography [e.g., Galloway and Rojstaczer, 1989] and fracture flow [e.g.,

Bower, 1983] have not been considered. While the model is simple, the result of the analysis is robust in that vertical flow may have occurred in the Arbuckle groundwater system. It calls into attention the potential problem for continued injection of large quantities of wastewater into this aquifer. It also shows that tidal detection of groundwater leakage can be useful for continuous monitoring the safety of groundwater source, the seepage from nuclear waste repository and the outflow of wastewater during hydrofracking. The method may also be used for the detection of earthquake-induced groundwater leakage [e.g., Wang, et al., 2016] by comparing water-level responses to Earth tides before and after an earthquake.

**Acknowledgement.** Work was partly supported by NSF grant EAR1344424 to C.Y.W.. We thank Michael Manga, Devin Galloway, the Associate Editor of WRR and three anomalous reviewers for reviewing the manuscript and providing constructive comments, Matthew Weingarten for discussion on the injection of wastewater in northeastern Oklahoma, and Lee-Ping Wang for helping with computation. All of the water level data presented in this paper are available on the World Wide Web through the National Water Information Service at https://waterdata.usgs.gov/nwis/uv/?site_no=364337096315401.

**References**

Agnew, D. C. (2012), SPOTL: Some programs for ocean-tide loading, SIO technical report, Scripps Institution of Oceanography, UC San Diego, Calif. [Available at http://escholarship.org/uc/sio_techreport.]

Allègre, V., E.E. Brodsky, L. Xue, S.M. Nale, B.L. Parker, and J.A. Cherry (2016), Using earth-tide induced water pressure changes to measure in situ permeability: A

comparison with long-term pumping tests, Water Resour. Res., 52, 3113–3126, doi:10.1002/2015WR017346.

Barbour, A.J., J.H. Norbeck, and J.L. Rubinstein (2017), The effects of varying injection rates in Osage County, Oklahoma, on the 2016 Mw 5.8 Pawnee earthquake, Seism. Res. Lett., 88, 1040-1053, doi: 10.1785/0220170003.

Bower D. R. (1983), Bedrock fracture parameters from the interpretation of well tides, J. Geophys. Res., 88, 5025–5035, doi:10.1029/JB088iB06p05025.

Cardiff, M., M. Barrash, and P.K. Kitanitis (2013), Hydraulic conductivity imaging from 3D transient hydraulic tomography at several pumping/observation densities, Water Resour. Res., 49, 7311-7326, doi:10.1002/wrcr.20519.

Christenson, S., N. I. Osborn, C. R. Neel, J. R. Faith, C. D. Blome, J. Puckette, and M.P. Pantea (2011). Hydrogeology and simulation of groundwater flow in the Arbuckle-Simpson aquifer, south-central Oklahoma, U.S. Geol. Surv. Sci. Invest. Rept. 2011-5029.

Cheng, A. H.-D. (1994), Reply to "Comment on 'Multilayered leaky aquifer systems: 1. Pumping well solutions' by A. H.-D. Cheng and O. K. Morohunfola'', Water Resour. Res., 30(11), 3231, doi:10.1029/94WR01803.

Cheng, A. H.-D., and O. K. Morohunfola (1993), Multilayered leaky aquifer systems: Pumping well solutions, Water Resour. Res., 29(8), 2787–2800, doi:10.1029/93WR00768.

Cihan, A., Q. Zhou, and J.T. Birkholzer (2011) Analytical solutions for pressure perturbation and fluid leakage through aquitards and wells in multilayered-aquifer systems, Water Resour. Res., 47, W10504,

doi:10.1029/2011WR010721.

Crank, J. (1956), *The Mathemetics for Diffusion*, Oxford University Press, Oxford.

Darrah, T. H., A. Vengosh, R. B. Jackson, N. W. Warner, and R. J. Poreda (2014), Noble gases identify the mechanisms of fugitive gas contamination in drinking-water wells overlying the Marcellus and Barnett shales, Proc. Natl. Acad. Sci. U.S.A., 111, 14,076–14,081.

Doan, M. L., E. E. Brodsky, R. Prioul, and C. Signier (2006), Tidal analysis of borehole pressure: A tutorial, Schlumberger-Doll Research Report, Cambridge, Mass. [Available at https://isterre.fr/IMG/pdf/tidal_tutorial_SDR.pdf.]

Edwards, P.J., L.L. Tracy, and W.K. Wilson (2011), Chloride concentration gradients in tank-stored hydraulic fracturing fluids following flowback, U.S. Dept. Agriculture, Research Paper-NRS-14.

Elkhoury, J. E., E. E. Brodsky, and D. C. Agnew (2006), Seismic waves increase permeability, Nature, 411, 1135–1138.

Ellsworth, W.L. (2013), Injection-induced earthquakes, Science, 1225942 -225943.

Frohlich, C. (2012), Two-year survey comparing earthquake activity and injection-well locations in the Barnett Shale, Texas, Proc. Natl. Acad. Sci. U.S.A., 109, 13,934–13,938.

Galloway, D.L., and Rojstaczer, S.A. (1989), Inferences about formation elastic and fluid flow properties from the frequency response of water levels to atmospheric loads and earth tides: 4th Canadian/American Conference on Hydrogeology:  Fluid Flow, Heat Transfer and Mass Transport in Fractured Rocks, Banff, Alberta, Canada, June 21-24, 1988, pp 100-113.

Hantush, M.S., and C.E. Jacob (1955), Non-steady Green's functions for an infinite strip of leaky aquifers, Transactions, American Geophysical Union, 36, 101.

Hantush, M. S. (1960), Modification of the theory of leaky aquifers, J. Geophys. Res., 65(11), 3713–3725, doi:10.1029/JZ065i011p03713.

Hemker, C. J. (1985), Transient well flow in leaky multiple-aquifer systems, J. Hydrol., 81, 111–126.

Hemker, C. J., and C. Maas (1994), Comment on "Multilayered leaky aquifer systems: 1. Pumping well solutions" by A. H.-D. Cheng, and O. K. Morohunfola, Water Resour. Res., 30(11), 3229–3230, doi:10.1029/94WR01802.

Hornbach, M.J., et al. (2015), Causal factors for seismicity near Azle, Texas, Nature Communications, doi: 10.1038/ncomms7728.

Hsieh, P., J. Bredehoeft, and J. Farr (1987), Determination of aquifer permeability from earthtide analysis, Water Resour. Res., 23, 1824–1832.

Ingebritsen, S. E., W. E. Sanford, and C. E. Neuzil (2006), *Groundwater in Geologic Processes*, 2nd ed., Cambridge Univ. Press, N. Y.

Johnson, K.S. (2008), Geologic history of Oklahoma, Oklahoma Geological Survey, Educational Publication 9.

Keranen, K.M. et al. (2013), Potentially induced earthquakes in Oklahoma, USA: Links between wastewater injection and the 2011 Mw 5.7 earthquake sequence, Geology, doi: 10.1130/G34045.1.

Keranen, K. M., M. Weingarten, G. A. Abers, B. A. Bekins, S. Ge (2014), Sharp increase in central Oklahoma seismicity since 2008 induced by massive wastewater injection, Science, 345, 448-451, doi: 10.1126/science.1255802.

Kroll, K.A., E.S. Cochran, and K.E. Murray (2017), Poroelastic Properties of the Arbuckle Group in Oklahoma derived from well fluid level response to the 3 September 2016 Mw5.8 Pawnee and 7 November 2016 Mw5.0 Cushing Earthquakes, Seism. Res. Lett., 88, doi: 10.1785/0220160228.

Lee, T.-C. (1999), *Applied Mathematics in Hydrogeology*, Lewis Publishers, N. Y.

Li, X.-M., B. Zhao, Z. Wang, M. Xie, J. Song, L.D. Nghiem, T. He, C. Yang, C. Li and G. Chen (2014), Water reclamation from shale gas drilling flow-back fluid using a novel forward osmosis–vacuum membrane distillation hybrid system, Water Science & Technology, 69.5, doi: 10.2166/wst.2014.003.

Maas, C. (1987a), Groundwater flow to a well in a layered porous medium, 1. Steady flow, Water Resour. Res., 23(8), 1675–1681, doi:10.1029/WR023i008p01675.

Maas, C. (1987b), Groundwater flow to a well in a layered porous medium, 2. Nonsteady multiple-aquifer flow, Water Resour. Res., 23(8), 1683–1688, doi:10.1029/WR023i008p01683.

McGarr, A. et al. (2015), Coping with earthquakes induced by fluid injection, Science, 347, 830-810, doi: 10.1126/science.aaa0494.

McNamara, D. E., H. M. Benz, R. B. Herrmann, E. A. Bergman, P. Earle, A. Holland, R. Baldwin, and A. Gassner (2015). Earthquake hypo- centers and focal mechanisms in central Oklahoma reveal a complex system of reactivated subsurface strike-slip faulting, Geophys. Res. Lett. 42, no. 8, 2742–2749, doi: 10.1002/2014gl062730.

Morgan, B. C. and K. E. Murray (2015), Characterizing small-scale permeability of the Arbuckle Group, Oklahoma. Open File Report OF2-2015, 1-12. National

Archives and Records Administration, 1989.

Rahi, K., and T. Halihan (2009). Estimating Selected Hydraulic Parameters of the Arbuckle-Simpson Aquifer from the Analysis of Naturally-Induced Stresses, Report, Oklahoma Water Resource Board.31. Wang, H.F., 2000, *Theory of Linear Poroelasticity*, 287 pp, Princeton: Princeton University Press.

Roeloffs, A. (1996), Poroelastic techniques in the study of earthquake-related hydrology phenomenon, Advances in Geophysics, 37:135–195, 1996.

Schoenball, M., & Ellsworth, W. L. (2017), A systematic assessment of the spatio-temporal evolution of fault activation through induced seismicity in Oklahoma and southern Kansas, J. Geophys. Res.: Solid Earth, 122, 10,189–10,206. https://doi.org/10.1002/2017JB014850

Sun, A.Y., J. Lu and S. Hovorka (2015), A harmonic pulse testing method for leakage detection in deep subsurface storage formation, Water Resour. Res., 47, 4263-4281, doi.org/10.1002/2014WR016567.

Tamura, Y., T. Sato, M. Ooe, and M. Ishiguro (1991), A procedure for tidal analysis 73 17 with a Bayesian information criterion, Geophys. J. Int., 104, 507–516.

U.S. Environmental Protection Agency (2016), Hydraulic Fracturing for Oil and Gas: Impacts from the Hydraulic Fracturing Water Cycle on Drinking Water Resources in the United States.

Veling, E. J. M., and C. Maas (2009), Strategy for solving semi-analytically three-dimensional transient flow in a coupled N-layered aquifer system, J. Eng. Math., 64, 145–161, doi:10.1007/s/10665-008-9256-9.

Vengosh, A., R. B. Jackson, N. Warner, T. Darrah, and A. Kondash (2014), A critical

review of the risks to water resources from unconventional shale gas development and hydraulic fracturing in the United States, Environ. Sci. Technol., 48, 8334–8348.

Vidic, R. D., S. L. Brantley, J. M. Vandenbossche, D. Yoxtheimer, and J. D. Abad (2013), Impact of shale gas development on regional water quality, Science, 340, doi:10.1126/science.1235009.

Walsh, F.R., and M.D. Zoback (2015), Oklahoma's recent earthquakes and saltwater disposal, Science Advances, doi 10.1126/sciadv.1500195.

Wang, H.F. (2000), *Theory of Linear Poroelasticity with Applications to Geomechanics and Hydrogeology*, Princeton University Press, Princeton, New Jersey.

Wang, C.-Y., Liao, X., Wang, L.-P., Wang, C.-H., Manga, M., 2016. Large earthquakes create vertical permeability by breaching aquitards. Water Resource Research, 52, doi:10.1002/2016WR018893.

Wang, C.-Y., M. Manga, M. Shirzaei, M. Weingarten, and L.-P. Wang (2017), Induced seismicity in Oklahoma affects shallow groundwater, Seism. Res. Lett., 88, 956-962, doi: 10.1785/0220170017.

Weingarten, M., et al. (2015), High-rate injection is associated with the increase in U.S. mid-continent seismicity, Science, 348, 1336-1340.

Xue, L., E.E. Brodsky, J. Erskine, P.M. Fulton, and R. Carter (2016), A permeability and compliance contrast measured hydrogeologically on the San Andreas Fault, Geochem. Geophys. Geosyst., 17, doi:10.1002/2015GC006167.

Yeck, W. L., G. P. Hayes, D. E. McNamara, J. L. Rubinstein, W. D. Barn- hart, P. S. Earle, and H. M. Benz (2016). Oklahoma experiences largest earthquake during ongoing regional wastewater injection hazard mitigation efforts, Geophys. Res. Lett. 44, no. 2, 711–717, doi: 10.1002/2016GL071685.

Yeck, W. L., M. Weingarten, H. M. Benz, D. E. McNamara, E. A. Berg- man, R. B. Herrmann, J. L. Rubinstein, and P. S. Earle (2016). Far- field pressurization likely caused one of the largest injection induced earthquakes by reactivating a large preexisting basement fault struc- ture, Geophys. Res. Lett. 43, no. 19, 10,198–10,207, doi: 10.1002/ 2016gl070861.

Zhang, Y., M. Person, J. Rupp, and 9 more authors (2013), Hydrogeologic controls on induced seismicity in crystalline basement rocks due to fluid injection into basal reservoirs, Groundwater, 51, 525-538.

**Supporting Information**

This Supporting Information consists of five parts: 1. Notations and definitions, 2. Verification of equation (15) by independent derivation using Laplace transform, 3. Vertical flow in the leaky aquifer model, and 4. Estimate of aquifer property based on purely unconfined aquifer model.

1. **Notations and definitions**

$h$: hydraulic head above a common reference.

$h_\infty$: hydraulic head away from the well above a common reference.

$h_w$: water level in the well

$\Delta h$: $h - h_\infty$

$\varepsilon$: the tidal oscillating volumetric strain of the aquifer

$h_{\infty,o}$: compex amplitude of hydraulic head away from the well

$h_{w,o}$: complex amplitude of water level in well

$\Delta h_o$: complex amplitude of $\Delta h$

$\varepsilon_o$: amplitude of $\varepsilon$

$T, S$: transmissivity and storativity of aquifer

$K'$: hydraulic conductivity of the semi-confining aquitard

$b'$: total thickness of the semi-confining aquitard

$B$: Skempton's coefficient

$K_u$: undrained bulk modulus

$r_w$: radius of the screened portion of the well

$r_c$: inner radius of well casing

## 2. Verification of equation (15) by independent derivation using Laplace transform

Let us suppose that a uniform dilatational stress $\sigma(t)$ is applied on a homogeneous isotropic aquifer, overlaid by a leaky aquitard. It induces a difference in hydraulic head in this aquifer, denoted $h(r,t)$, and a change in water level in a well, $h_w(t)$. The system is controlled by the set of equations:

$$\frac{\partial^2 h}{\partial r^2} + \frac{1}{r}\frac{\partial h}{\partial r} - \frac{K'}{Tb'}h + \frac{BS}{\rho g T}\frac{\partial \sigma}{\partial t} = \frac{S}{T}\frac{\partial h}{\partial t} \qquad (S2.1)$$

$$\rho g h(r \to \infty) = B\sigma \qquad (S2.2)$$

$$h(r = r_w) = h_w \qquad (S2.3)$$

$$2\pi r_w T \left.\frac{\partial h}{\partial r}\right|_{r=r_w} = \pi r_c^2 \frac{\partial h_w}{\partial t} \qquad (S2.4)$$

$S$ and $T$ are the storativity and the transmissivity of the aquifer. We also distinguish between the casing radius ($r_c$) in which the water level rises and the well radius ($r_w$) where the water flow enters the well. $B$ is Skempton's coefficient which relates the pore pressure $P = \rho g h$ variation to the load $\sigma$ applied to the porous media. $K'$ and $b'$ are the vertical hydraulic conductivity and the thickness of the aquitard.

If there were no well, the problem would be laterally invariant, so that all the $r$-derivatives are null. The change in hydraulic head would express itself after a Laplace transformation as $h_\infty e^{pt}$ (for sinusoidal signals, $p = i\omega$). Equation (S2.1) simplifies into:

$$-\frac{K'}{b'}h_\infty + \frac{BS}{\rho g}p\sigma = S\,p\,h_\infty \tag{S2.5}$$

This is the same as equation (6) in the text.

With the well effect, we consider the quantity $s = h - h_\infty$, The system is controlled by the set of equations

$$\frac{\partial^2 s}{\partial r^2} + \frac{1}{r}\frac{\partial s}{\partial r} - \frac{K'}{Tb'}s = \frac{S}{T}\frac{\partial s}{\partial t} \tag{S2.6}$$

$$\rho g\, s(r \to \infty) = 0 \tag{S2.7}$$

$$s(r = r_w) + h_\infty(t) = h_w \tag{S2.8}$$

$$2\pi r_w T \left.\frac{\partial s}{\partial r}\right|_{r=r_w} = \pi r_c^2 \frac{\partial h_w}{\partial t} \tag{S2.9}$$

We eliminated the loading term from the partial differential condition to confine it to the boundary conditions. We then apply Laplace transform on (S2.6) to (S2.9)

$$\frac{\partial^2 s(r,p)}{\partial r^2} + \frac{1}{r}\frac{\partial s(r,p)}{\partial r} - \frac{K'}{Tb'}ps = \frac{S}{T}ps \tag{S2.10}$$

$$s(r \to \infty, p) = 0 \tag{S2.11}$$

$$s(r = r_w, p) = h_w(p) - \frac{p}{p + \frac{K'}{b'S}}\frac{B}{\rho g}\sigma \tag{S2.12}$$

$$2\pi r_w T \left.\frac{\partial s(r,p)}{\partial r}\right|_{r=r_w} = p\pi r_c^2 h_w \tag{S2.13}$$

The solution to equation (S2.10) is $C_I I_0(qr) + C_K K_0(qr)$, with $q = \sqrt{p\frac{S}{T} + \frac{K'}{Tb'}}$.

Equation (S2.10) asserts than $C_I = 0$. We have then two unknowns to solve, $C_K$ and $h_w(p)$ with the two equations (S2.12) and (S2.13). We then get

$$h_w = \cfrac{1}{1 + \cfrac{r_c^2}{r_w^2} \cfrac{\alpha}{2S} \cfrac{K_0(\alpha)}{K_1(\alpha)} + \cfrac{K'}{pSb'}} \cfrac{B\sigma}{\rho g} \qquad (S2.14)$$

where $\alpha = qr = r_w \sqrt{p\dfrac{S}{T} + \dfrac{K'}{Tb'}}$.

With $\sigma$ replaced by $BK_u$, $p$ by $i\omega$, and $\alpha$ by the above expression, (S2.14) becomes

$$h_w = \cfrac{1}{1 + \cfrac{r_c^2}{r_w^2} \cfrac{\beta r_w}{2S} \cfrac{K_0(\beta r_w)}{K_1(\beta r_w)} + \cfrac{K'}{i\omega Sb'}} \cfrac{BK_u \epsilon}{\rho g} = \cfrac{1}{1 + \cfrac{K'}{i\omega Sb'} + \cfrac{r_c^2}{r_w^2} \cfrac{\beta r_w}{2S} \cfrac{K_0(\beta r_w)}{K_1(\beta r_w)}} \cfrac{BK_u \epsilon}{\rho g} \qquad (S2.15)$$

Finally, by multiplying both the numerator and the denominator by

$$\cfrac{i\omega S}{i\omega S + \cfrac{K'}{b'}} = \cfrac{1}{1 + \cfrac{K'}{i\omega Sb'}} = \cfrac{i\omega S}{T\beta^2}, \text{ we get}$$

$$h_w = \cfrac{\cfrac{i\omega S}{i\omega S + \cfrac{K'}{b'}}}{1 + \cfrac{r_c^2}{\beta r_w} \cfrac{i\omega}{2T} \cfrac{K_0(\beta r_w)}{K_1(\beta r_w)}} \cfrac{BK_u \epsilon}{\rho g} \qquad (S2.16)$$

which is identical to equation 15 in the main text.

### 3. Vertical flow in the leaky aquifer model

As noted in the text, equation (15) cannot be reduced to the classical solution for a purely unconfined aquifer at $T = 0$ because of the linear approximation of the Darcy's law by $-K'h/b'$ for the vertical leakage and the implicit assumption that the aquitard is incompressible and has negligible storage. Nevertheless, purely vertical flow does occur in the present model away from the well, which is $-(K'/b')h_{\infty,o}$ where $h_{\infty,o}$ is given by equation (7). In order to compare this solution with that for purely unconfined aquifers, we define a dimensionless frequency [Galloway and Rojstaczer, 1989]

$$\varpi = \omega b'^2/D' \tag{S3.1}$$

where $D' = K'/S_s$ and $S_s = S/b'$. In terms of $\varpi$, we may express equation (7) as

$$\frac{\rho g h_{\infty,o}}{B K_u \varepsilon_o} = \frac{1+i(1/\varpi)}{1+(1/\varpi)^2}. \tag{S3.2}$$

For purely unconfined aquifers, a similar ratio may be expressed as [e.g., Wang, 2000]

$$\frac{\rho g h_o}{B K_u \varepsilon_o} = 1 - \exp\left[(1+i)\left(\frac{z}{\sqrt{\frac{2D}{\omega}}}\right)\right]. \tag{S3.3}$$

where $z$ is the depth of the screening interval of a cased well. Identifying $\frac{z}{\sqrt{2}}$ with $b'$ we express the above equation in terms of $\varpi$ as

$$\frac{\rho g h_o}{B K_u \varepsilon_o} = 1 - exp[(1+i)\sqrt{\varpi}]. \tag{S3.4}$$

Figure S1 shows a plot of the predicted response of amplitude ratio and phase shift to the M2 tide of equation (S3.2) and of the purely unconfined aquifer model (S3.4) [Roeloffs, 1996; Wang, 2000] as functions of log $\varpi$. It shows that the two models are in good agreement in their predicted amplitude ratios. It also shows that the two models agree in the sign of the predicted phase shift; but the magnitude of the predicted phase shift by equation (7) is greater than that predicted by the unconfined model by a factor of 2. This difference reflects the fact noted in the main text that, while the classical unconfined aquifer model is specifically that of a half space, the leaky aquifer model developed here is for an aquifer of finite thickness and confined below.

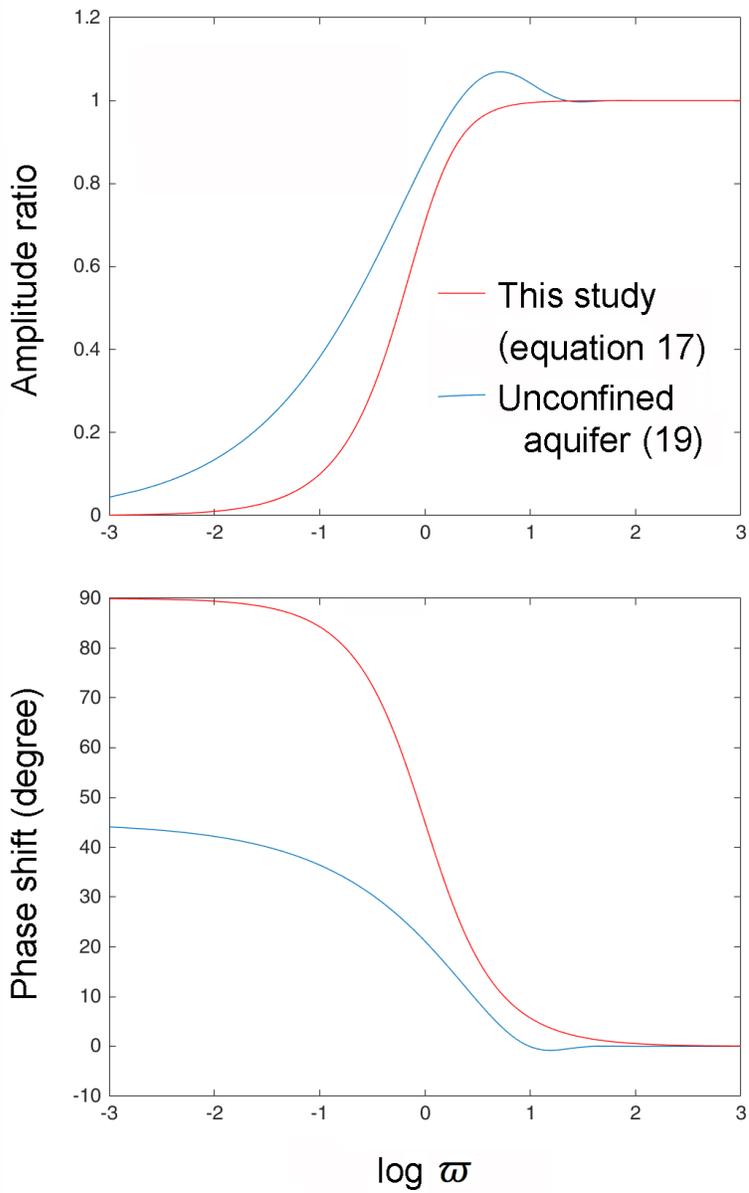

*Figure S1. Predicted response of amplitude ratio and phase shift to the M2 tide of purely vertical flow in the present model (red) and the unconfined aquifer model (blue), plotted as functions of the logarithm of the dimensionless frequency $\varpi$.*

4. **Estimate of aquifer property based on purely unconfined aquifer model**

For purely unconfined aquifers, pore pressure response to solid tides may be expressed as [Roeloffs, 1996; Wang, 2000]:

$$P(z) = \gamma\sigma_0 \left\{1 - \exp\left[-z\sqrt{\frac{\omega}{2D}}\right]\exp\left[-iz\sqrt{\omega/2D}\right]\right\} \tag{S4.1}$$

where, $\gamma$ is the loading efficiency, $\sigma_o$ is the amplitude of the imposed tidal forcing, $z$ is the depth from the water table, $\omega$ is the angular frequency of the imposed tidal forcing, $D$ is the hydraulic diffusivity. The corresponding phase shift is given by

$$\eta = \arg\left\{\frac{\exp\left(-\frac{z}{\delta}\right)\sin\frac{z}{\delta}}{1-\exp\left(-\frac{z}{\delta}\right)\sin\frac{z}{\delta}}\right\} \tag{S4.2}$$

where $z$ is the depth of the well and $\delta = \sqrt{\frac{2D}{\omega}}$ is the characteristic diffusion length.

For the USGS deep Oklahoma monitoring well, $z \sim 960$ m, $\omega/2\pi = 1.9324$ cpd, and the observed phase shift for M2 is $\sim 12.5°$. The diffusivity inverted by a grid search based on equation (S4.2) is $D \sim 19$ m$^2$/s. This corresponds to an average hydraulic conductivity of $2\times 10^{-5}$ m/s across a depth range of 960 m, assuming an average specific storage of $10^{-6}$ m$^{-1}$. The hydraulic conductivity so estimated seems too high for the depth range of the well. We argue, as we did in the text, that a better model for the interpretation of the tidal response of water level in the USGS deep well is the leaky aquifer model because both geologic studies [e.g., Johnson, 2008] and well logs (Figure 5) show that the Arbuckle aquifer is overlain by layers of younger rocks that includes a basal shale. Thus the positive phase shift observed at the USGS well is better interpreted as indicating leakage of the confinement, where both vertical and horizontal flow occurs and contributes to the measured phase response, rather than as purely unconfined flow where only vertical flow occurs.